\documentclass[useAMS,usenatbib,usegraphicx]{mn2e}
\usepackage{times}

\title[The metallicity dependence of the LGRB rate]{The metallicity
  dependence of the long-duration GRB rate from host galaxy luminosities}

\author[C. Wolf and Ph.\ Podsiadlowski]{Christian Wolf and Philipp
 Podsiadlowski \smallskip \smallskip \\
Dept.\ of Physics, University of Oxford, Keble Road, Oxford, OX1 3RH,
U.K., email: cwolf@astro.ox.ac.uk, podsi@astro.ox.ac.uk}

\begin{document}
\date{submitted}
\maketitle

\begin{abstract}
We investigate the difference between the host galaxy properties of
core-collapse supernovae and long-duration gamma-ray bursts (LGRBs),
and quantify a possible metallicity dependence of the efficiency of
producing LGRBs. We use a sample of 16 CC SNe and 16 LGRBs from
\citet{F06} which have similar redshift distributions to eliminate
galaxy evolution biases. We make a forward prediction of their host 
galaxy luminosity distributions from the overall cosmic metallicity
distribution of star formation. The latter is based on luminosity 
functions, star-formation rates (SFR) and luminosity--metallicity 
($L$--$Z$) relations of galaxies. This approach is supported by the 
finding that LGRB hosts follow the $L$--$Z$-relations of star-forming
galaxies. We then compare predictions for metallicity-dependent 
event efficiencies with the observed host data. We find that 
UV-based SFR estimates predict the hosts distribution of CC SNe 
perfectly well in metallicity-independent form. In contrast, LGRB 
hosts are on average fainter by one magnitude, almost as faint
as the Large Magellanic Cloud. Assuming this is a metallicity effect, 
the present data are insufficient to discriminate between a sharp 
cutoff and a soft decrease in efficiency towards higher metallicity. 
For a sharp cut-off, however, we find a best value for the cutoff 
metallicity, as reflected in the oxygen abundance, $12+\log {\rm 
(O/H)}_{\rm lim}\simeq 8.7\pm 0.3$ at 95\% confidence including 
systematic uncertainties on the calibration of \citet{KK04}. This 
value is somewhat lower than the traditionally quoted value for 
the Sun, but is comparable to the revised solar oxygen abundance 
\citep{A05}. LGRB models that require sharp metallicity cutoffs 
well below $\sim 1/2$ the revised solar metallicity appear to be 
effectively ruled out, as they would require fainter LGRB hosts than 
are observed. We also discuss the likely implications of the still
ongoing metallicity 'calibration debate'.
\end{abstract}


\section{Introduction}

Since the first optical afterglow to a gamma-ray burst was identified
\citep[ GRB 970228]{vP97}, the restframe properties of about 50 host
galaxies of long-duration gamma-ray bursts (LGRBs) have been
identified. Already with a small sample it became clear that the hosts 
are preferentially faint blue irregular galaxies \citep{F99,LeF03,CHG04}. 
This trend was recently confirmed by a comparison between hosts of LGRBs 
and hosts of core-collapse supernovae (CC SNe). CC SNe were found in 
late-type galaxies of all morphologies and luminosities, in contrast 
to LGRB hosts, which were still found to be almost exclusively irregular
and on average fainter \citep[ hereafter F06]{F06}.

Since both CC SNe and LGRBs are considered explosions of massive young
stars with ages below $\sim 30$~Myr and probably below the typical
timescale of star formation in most galaxies, there is {\em apriori\/}
no reason to expect a considerable difference in their host
galaxies. CC SNe should be direct and unbiased tracers of very recent
star formation, although complications may arise from dust extinction
intrinsic to the host. Furthermore, we expect a dependence of the mass 
threshold for core collapse/black-hole formation on metallicity \citep{
H95,H03,ET04} and binary evolution effects \citep{B99,B01,Pf02,P04b}.

The suggested explanation for the host preference of LGRBs is that
they arise from low-metallicity stellar populations which are
predominantly found in low-luminosity galaxies according to the
luminosity--metallicity relation \citep[e.g][ hereafter KK04]{KK04}. 
Also, in
several progenitor models of GRBs \citep[in particular `collapsar' 
models;][]{MW99}, wind mass loss plays an important role in slowing down
the rotation of massive stellar cores. Since lower metallicity leads
to weaker stellar winds and hence less angular-momentum loss, the 
stars are more likely to retain rapidly rotating cores at the time of 
the explosion, as required in the collapsar model.

%
%
%
%

LGRBs discovered by the {\it Swift} mission have a median redshift of 2.75 
and apparently have a distribution expected for an unbiased tracer of star 
formation \citep{J06}. If both the cosmic star formation history and the 
cosmic evolution of metallicity in cold star-forming gas were known with 
high accuracy, this redshift distribution could provide clues about the 
metallicity dependence of the LGRB rate. However, the LGRBs discovered by 
BeppoSAX have a median redshift near $z\simeq 1$, and predictions for the
completeness function are very difficult to make, owing to the complexity 
of GRB trigger algorithms and detector intricacies. Even with a much larger 
sample, the redshift distribution of GRBs might be affected by systematic
uncertainties such that no strong constraints on metallicity can be found. 

Instead, first spectroscopic measurements of the metallicity of LGRB hosts
at redshifts $0.4<z<1$ from the GHost Studies \citep{Sav06} indicate that
LGRB hosts follow the luminosity--metallicity relation of star-forming 
galaxies. A preference for low metallicity would then require LGRBs to 
favour less luminous galaxies more than general star formation does. When
working with gas-phase metallicities of galaxies, we need to be aware of 
the {\it calibration debate} between two basic methods which result in
numbers that differ by a factor of $\sim 2$: spectra of low signal-to-noise
only facilitate an indirect strong-line method (Pagel's $R_{23}$ indicator,
1979) to measure metallicity, which is calibrated against photoionization 
models. In contrast, high-S/N spectra allow for a direct measurement of
metallicity taking into account the electron temperature measured from
weak auroral lines \citep[see e.g. ][]{K03}. The $T_e$-based methods find
typically 0.2--0.5 dex lower metallicities, but have been critized to be 
biased by not taking inhomogeneities in the temperature into account.

In this paper, we wish to derive constraints on the metallicity dependence
of LGRB progenitors by comparing forward predictions of host luminosity
distributions with the observed host sample. We apply a minor modification 
to the F06 samples by restricting them to identical redshift ranges $z=
[0.2,1]$ (see Sect.~2). Our quantitative predictions of the host luminosity 
distribution is based on luminosity functions and star-formation rates of 
galaxies and their luminosity--metallicity relations (see Sect.~3). Here,
we need to make sure that all ingredients are on the exact same metallicity 
calibrator and opt for the strong-line method on the KK04 calibrator for
consistency. We then fold in different prescriptions for the metallicity 
dependence of the LGRB event rate and constrain these by comparing our 
predictions to the observed sample in Sect.~4. In Sect.~5 we discuss the 
results and the implications of the calibration debate. Throughout the 
paper, we use Vega magnitudes and the cosmological parameters $(\Omega_{
\rm m},\Omega_\Lambda) =(0.3,0.7)$ and $H_0=h_{70}\times 70$~km/(s~Mpc). 
We use the revised solar oxygen abundance of $12+\log {\rm (O/H)}=8.66$ 
\citep[see also \citet{ALA01}]{A05}.

\section{Data}

This work is based on the host galaxy data of CC SNe and LGRBs
collected by F06. The CC SNe were all discovered by the Hubble Higher
Redshift Supernova Search \citep{S04} in the two HST GOODS fields
\citep{G04}. This search was only sensitive to CC SNe at $z\le 1$ and
may already be incomplete at the upper end of this redshift range
\citep{Da04}. The LGRBs were taken by F06 from public HST data of
host galaxy observations, where the list of objects extends to
redshifts above 4.

At redshifts $z<0.2$, the HST GOODS survey has too little volume to
detect CC SNe efficiently. The same restriction applies to samples of
LGRBs with typical gamma-ray luminosities. However, the GRB list
contains a few very nearby objects with extremely low gamma-ray
luminosities, which may be viewed at larger angles or be intrinsically
faint events, and possibly an altogether different kind of explosion
\citep[see the ensuing discussion on GRB 060218, e.g.][]{Sod06}. In
order to exclude these peculiar GRB events, which we could only see at
redshifts $z\la 0.1$, we restrict the existing CC SN and LGRB data to
a subsample in the redshift range $z=[0.2,1.0]$. We note, that this is 
similar to but more conservative than the comparison considered by F06, 
which includes LGRBs up to $z=1.2$.

As a result, we keep the full sample of 16 CC SNe with a mean redshift of
0.63, while the LGRB sample is restricted to 16 events with a mean redshift 
of 0.69, so they are on average from an extremely similar cosmic epoch.

\section{Modelling the host luminosity distribution}

We wish to model restframe $V$-band luminosity distributions for the host 
galaxies of both types of explosive events in order to compare them with
the observed distribution from the restricted F06 sample. For this purpose
we ultimately need a distribution of star-formation density over $V$-band 
host luminosity and potentially a metallicity-dependent efficiency function.

\subsection{Galaxy luminosity function and star-formation density}

\citet[ hereafter W05]{W05} measured the restframe UV (280~nm)
luminosity function (LF) of galaxies from a sample of almost 1500
galaxies at a mean redshift of $z\simeq 0.70$. Based on morphological
classifications from the GEMS survey \citep{R04,C06}, the LFs were
also split by galaxy type. The mean redshift of these LFs coincides
with that of the host galaxy samples for the CC SNe and LGRBs
considered here. Hence, they can serve as an unbiased description of
the general galaxy population at the redshift of the events. The LF 
$\phi(L_{280})$, is parameterized as a Schechter function

\begin{equation}\label{LF}
  \phi(L_{280}) dL = \phi^* (L/L^*)^\alpha \exp{-L/L^*} dL	~.
\end{equation}
We chose the restframe UV luminosity at 280~nm, $L_{280}$, since it can be 
used as a proxy for the unobscured star-formation rate using the relation of
\citet{K98}

\begin{equation}
	{\rm SFR} (M_\odot {\rm yr}^{-1}) = 1.4\times 10^{-28} L_{280} /
	({\rm erg~s}^{-1}{\rm ~Hz}^{-1})	~ .
\end{equation} 
Evidently, the total star-formation rate of a galaxy depends also on the 
amount of obscured star formation, which could be measured from the thermal 
FIR dust emission or crudely estimated using UV spectral slopes.

However, both CC SNe and LGRB afterglows are subject to host galaxy
extinction as well, just as expected for events occuring in very young
stellar populations.  In fact, star-bursting galaxies do not only have
an above average proportion of obscured star formation, but also an
increased rate of obscured, optically invisible CC SNe, which can be
found by NIR monitoring \citep{Man03}. Such SNe are absent in the
sample used here, which is based on a restframe UV-optical
search. Present data for LGRB afterglows show no evidence for high
levels of dust obscuration, while a fraction of optically dark LGRBs
may still be due to high dust extinction in the host galaxy
\citep{Kl03}. However, the host galaxies of known LGRBs show no
significant signs of dust-obscured star formation \citep{L06}.

Hence, we conclude that a comparison of UV-based star formation
measures with UV-optical detection rates of stellar explosions is
fair. It would be next to impossible to attempt an accurate extinction
correction, which would need to take into account changes in the
average extinction with progenitor lifetime as well as detailed
detection efficiencies for events of different extinction levels.

For our model, we first calculate the distribution of UV luminosity density 
$j_{280}$ cumulative over UV host luminosity $L_{280}$ using the integral

\begin{equation}\label{HDflat}
  j_{280} (L_{280,\rm lim}) = \int_{L_{280,\rm lim}}^{\infty} 
	 L_{280} \, \phi(L_{280}) \, dL_{280} ~.
\end{equation}
Given the Kennicutt law, this quantity is an estimate of the 
unobscured star-formation density in hosts of $L_{280}>L_{280,\rm lim}$. 
Hence, it also describes the host distribution of any explosive event which 
is proportional to the star-formation rate.  As an example, CC SNe are
broadly expected to be an unbiased tracer of star formation, and should 
have a host distribution described by (\ref{HDflat}).  

In a second step, we transform the UV luminosity axis $L_{280}$ into $L_V$
to match the given luminosity data of the host sample. For this purpose, we
need to know the $(M_{280}-M_V)$-colours of star-forming galaxies. These 
data have been obtained by the COMBO-17 survey and are publicly available 
for the Chandra Deep Field South \citep{W04} covering the entire GEMS field.
The galaxy sample used for the above LFs can be split into a red sequence of
mostly passively evolving galaxies and a blue cloud of mostly star-forming
galaxies. For star-forming galaxies at $z\simeq 0.7$, the colour-magnitude 
relation is (see Fig.~5 of W05)

\begin{equation}
  M_{280}-M_V = - 0.11 (M_{280}-5\log h_{70} +20)   ~.
\end{equation} 
Hence, the lower integration limit in the calculation of the luminosity density 
can be substituted as 

\begin{equation}
  M_{V,{\rm lim}}=M_{280,{\rm lim}}+0.11 (M_{280,{\rm lim}}-5\log h_{70} +20) ~.
\end{equation} 
This is equivalent to stretching the scale of the luminosity axis by 11\% and
holding it fixed at $M_V=M_{280}=-20$, where the average colour of star-forming
galaxies is zero. 

\begin{table}
\caption{UV luminosity functions of galaxies at $z\simeq 0.7$ from COMBO-17
and GEMS (transformed from $h$ to $h_{70}$).\label{LFs}}
\begin{tabular}{lcc}
\hline \noalign{\smallskip} 
sample			&  $M^*_{280}-5\log h_{70}$	& $\alpha$  \\
\noalign{\smallskip} \hline \noalign{\smallskip}
all galaxies		& $-20.45\pm0.16$ 	& $-0.75\pm0.20$ \\
irregular galaxies	& $-20.11\pm0.17$ 	& $-1.31\pm0.20$ \\
\noalign{\smallskip} \hline
\end{tabular}
\end{table}

\begin{figure}
\centering
\includegraphics[clip,angle=270,width=0.97\hsize]{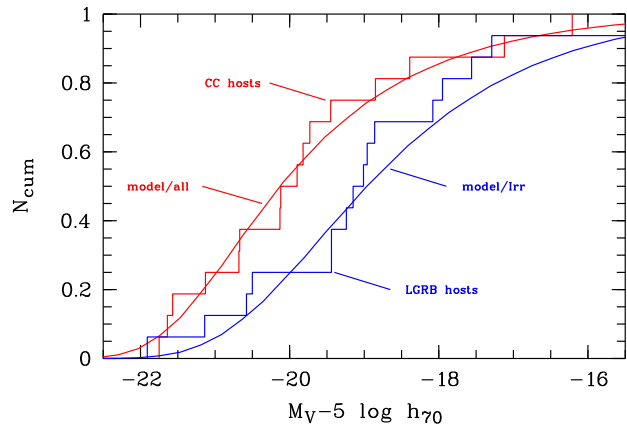}
\caption{Cumulative luminosity distributions of CC SN and LGRB host
galaxies (step functions), together with distributions for the
UV-derived star-formation density in all galaxies and in irregular
galaxies alone (curves). \label{toy}}
\end{figure}

\subsection{Predictions by galaxy type}

It has often been reported, most recently by F06, that CC SNe appear
in all types of star-forming galaxies, while the afterglows of LGRBs 
are almost exclusively found in morphologically irregular, faint blue
galaxies. Although galaxy morphology may not be a suitable physical
factor for the evolution of high-mass stars in a galaxy, it may be
correlated with physical factors: irregulars are mostly of
lower-luminosity and hence tend to be of lower metallicity, which may
well be the physical cause of the increased LGRB rate.

We now briefly try to confirm the picture suggested by F06, and test a
toy model in which CC SNe arise unbiased from (unobscured) star
formation in all galaxies while LGRBs arise unbiased from irregular
galaxies. We model the host luminosity distributions using the
parameters of the LFs obtained by W05 for irregular galaxies and {\it
all\/} galaxies (see Tab.~\ref{LFs}). Note that the model is only a
normalised cumulative luminosity distribution. Hence, it does not
constrain overall rates, and it also does not depend on the
normalisation $\phi^*$ of the LF, but only on the slope $\alpha$ of
the faint end and the location $M^*$ of the break in the LF.

This toy model is compared with the sample data in Fig.~\ref{toy}. We
find that CC SNe match the prediction extremely well and conclude that
they can be considered an unbiased proxy for star formation in all
galaxies within the limits of this small data set. Also, the
prediction for irregular galaxies matches the host data for LGRBs
reasonably well, although there is some deviation at low luminosities. 
Formally, both comparisons are entirely acceptable: a Kolmogorov-Smirnov 
test yields values of $D_{\rm KS}=0.12$ and $0.16$, respectively 
(corresponding to rejection probabilities of only 4\% and 22\%).

We note, that the LFs are well-determined at $M_V< -18.5$ but
extrapolated with a Schechter function at fainter magnitudes.  Any
errors in this extrapolation will translate into a simple rescaling of
the cumulative host prediction at $M_V$ brighter than $-18.5$, whereas
the fainter section would change its shape. We choose to ignore the
errors in the previous qualitative analysis and present a more
quantitative study, using metallicity as the physical factor controlling
the LGRB efficiency, in the following sections.

\begin{table}
\caption{$L$--$Z_{\rm O}$-relations at $z\simeq 0.7$ from Kobulnicky \&
  Kewley (2004) using a functional form of $12+\log {\rm(O/H)} = a +
  b\, M_B$.
\label{KK04fits}}
\begin{tabular}{lcc}
\hline \noalign{\smallskip}  
fit	&  $a$		&  slope $b$  \\
\noalign{\smallskip} \hline \noalign{\smallskip}
A	& $6.37$ 	& $-0.117$ \\
B (best)& $3.88$ 	& $-0.239$ \\
C	& $1.37$ 	& $-0.363$ \\
\noalign{\smallskip} \hline
\end{tabular}
\end{table}

\subsection{Luminosity--metallicity relations and dispersion in the 
  stellar population}

In this section, we refine our toy model by considering as progenitors
only sub-populations selected by metallicity. Our use of the term
metallicity denotes the abundance of oxygen in the ionised gas phase
as a tracer of metallicity in young high-mass stars. We use the O/H
ratio measured via the $R_{23}$-index on the KK04 calibration. This 
index is determined from flux ratios in the prominent ionized oxygen 
and H$\beta$ emission lines of star-forming galaxies, which are 
relatively easy to observe even in faint galaxies at higher redshifts. 
Of course, it would also be attractive to look at abundances of other 
elements. However, these are more challenging to measure, and we
believe that, with the present small sample of objects, we would not
be able to improve our study significantly. As an abbreviation we will
use the notation

\begin{equation}
  Z_{\rm O} = 12+\log {\rm(O/H)}	~.
\end{equation} 
Recent determinations of the luminosity--metallicity ($L$--$Z_{\rm
O}$) or the stellar mass--metallicity ($M_*$--$Z_{\rm O}$) relation 
include the work by \citet[ herafter KK04]{KK04} on the basis of
the GOODS survey, as well as \citet{Sav05} using the GDDS
survey. We opt for using the $L$--$Z_{\rm O}$-relation, which relates
more directly to our data and model. For star-forming galaxies at
$z\simeq 0.7$, the COMBO-17 data suggest on average $M_B \simeq M_V$,
and we use the KK04 relations without further modifications.  Their
relations are determined in redshift slices, and conveniently their
slice $z=[0.6,0.8]$ corresponds precisely to the redshift range in
which the LF was determined by W05. Altogether, three linear fits are
shown in their Fig.~11: a $Z_{\rm O} = f(M_B)$-fit (A), an $M_B =
f(Z_{\rm O})$-fit (C) and a bisector fit (B), which can all be
expressed as

\begin{equation}\label{Zfit}
  Z_{\rm fit} = 12+\log {\rm(O/H)} = a + b \times M_B	~.
\end{equation}
The bisector fit has a slope of $b=0.239$, between the two one-sided
fits, and is presumably the most realistic one of the three. We also
use the two one-sided fits to define a confidence interval for the
relation (see Tab.~\ref{KK04fits} for fit parameters).

A recent determination of the $L$--$Z_{\rm O}$-relation in local dwarf
galaxies suggests a slope of $b\simeq 0.30$ extending across the
luminosity range from $M_B=-11$ to $-19$ \citep{Lee06}. At fixed
luminosity, the metallicity has a $1\sigma$-scatter of 0.16~dex, the
same as that determined in a large sample of luminous galaxies from
SDSS \citep{Tre04}. The scatter of the mass--metallicity relation is
even less (0.10), and in the absence of any higher-redshift data we
assume that it is well-behaved also for dwarfs at $z\simeq 0.7$. In
any case, both the $L$--$Z_{\rm O}$-relation and the UV luminosity
function are just extrapolated below $M_V=-18.5$ in our analysis.

Below we first investigate the effects of a sharp metallicity cutoff
for the production of LGRBs in our model. If we ignored both the
scatter in the $L$--$Z_{\rm O}$-relation and the internal variations of
$Z_{\rm O}$ among the young stellar population within any given
galaxy, a sharp $Z_{\rm O}$-cutoff would cause a sharp $M_V$-cutoff in
the host galaxy sample. While we do not have perfect knowledge of the
overall $Z_{\rm O}$ scatter, it certainly needs to be taken into
account for the prediction of cumulative host luminosity distributions
in the presence of $Z_{\rm O}$-dependent event efficiencies.

Even for the same data, the scatter of the $L$--$Z_{\rm O}$-relation
for average galaxy metallicities varies with the slope of the
fit. Roughly, it is around $\pm 0.3$~dex end-to-end for a given
luminosity at $z\simeq 0.7$, consistent with a $1\sigma$-scatter of
0.16 at low redshift. As far as the $Z_{\rm O}$-spread of star-forming
gas within a galaxy is concerned, we lack useful knowledge at $z\simeq
0.7$ owing to the limited spatial resolution of spectroscopic data,
especially for smaller galaxies.

Instead, we look at the present-day Magellanic clouds for clues on the
$Z_{\rm O}$ variation within irregulars. The LMC appears to resemble a
typical LGRB host at $z\sim 0.7$ given its luminosity of $M_V\simeq
-18.5$. The LMC and SMC have measured metallicities in young stars and
the gas phase of 0.5 solar and 0.2 solar, respectively, and both have
internal 1$\sigma$-variations of $\la 0.2$~dex \citep{RB89}. Dispersions 
in Milky Way-size disk galaxies are not expected to be larger either, 
given efficient gas mixing.

In principle, we need to convolve the distribution of the internal
$Z_{\rm O}$ variation and the scatter in the $L$--$Z_{\rm O}$-relation
to obtain a realistic cosmic $p(Z_{\rm O}|M_V)$-distribution. But in
the absence of any accurate values for the former, we consider two
alternative scenarios for the combined metallicity variation by using 
the form

\begin{eqnarray}
  d_Z(Z_{\rm O},M_V) & = & \frac{Z_{\rm O}-Z_{\rm fit}(M_V)}{w}, \\
    p(Z_{\rm O}|M_V) & = & \left\{  \begin{array}{cl}
  			0 		& \mbox{~if $|d_Z|>1$}, \\
  			\cos d_Z	& \mbox{~if $|d_Z|\le 1$}.  
  \end{array} \right.
\end{eqnarray}
where $Z_{\rm fit}$ is taken from the $L$--$Z_{\rm O}$-relation (\ref{Zfit}) 
and $\pm w$ is the end-to-end spread in the variation applied at any fixed 
$M_V$. We opted to use a cosine function, because it is trivial to integrate 
analytically. Two other simple alternatives appear unrealistic, a box 
function because of its steep edges and a Gaussian because of its infinitely 
far-reaching wings.

Overall, we consider $w=0.4$~dex a realistic value for the combined
spread at fixed host luminosity, which corresponds to about $\pm 0.25$~dex 
for the FWHM of the distribution $p$. Just for illustration we
also explore an extreme value of $w=1$, which translates into an
extreme and unrealistic metallicity spread from ten-fold below to
ten-fold above the fit value at any given luminosity. This will allow 
even very luminous galaxies in our model to form a fraction of very 
low-$Z_{\rm O}$ stars.

\begin{figure*}
\centering
\includegraphics[clip,angle=270,width=0.47\hsize]{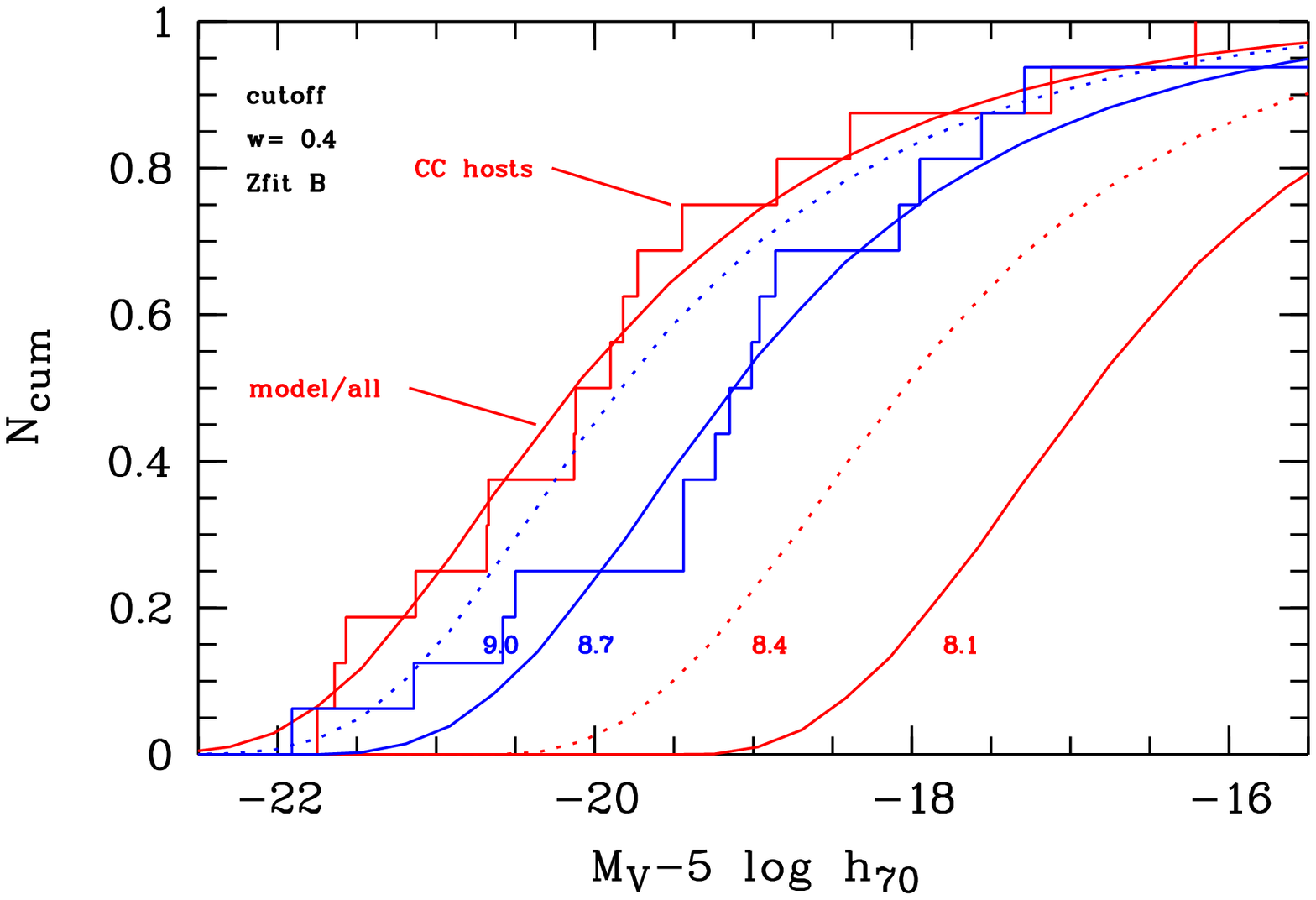}
 \hspace{5mm}
\includegraphics[clip,angle=270,width=0.47\hsize]{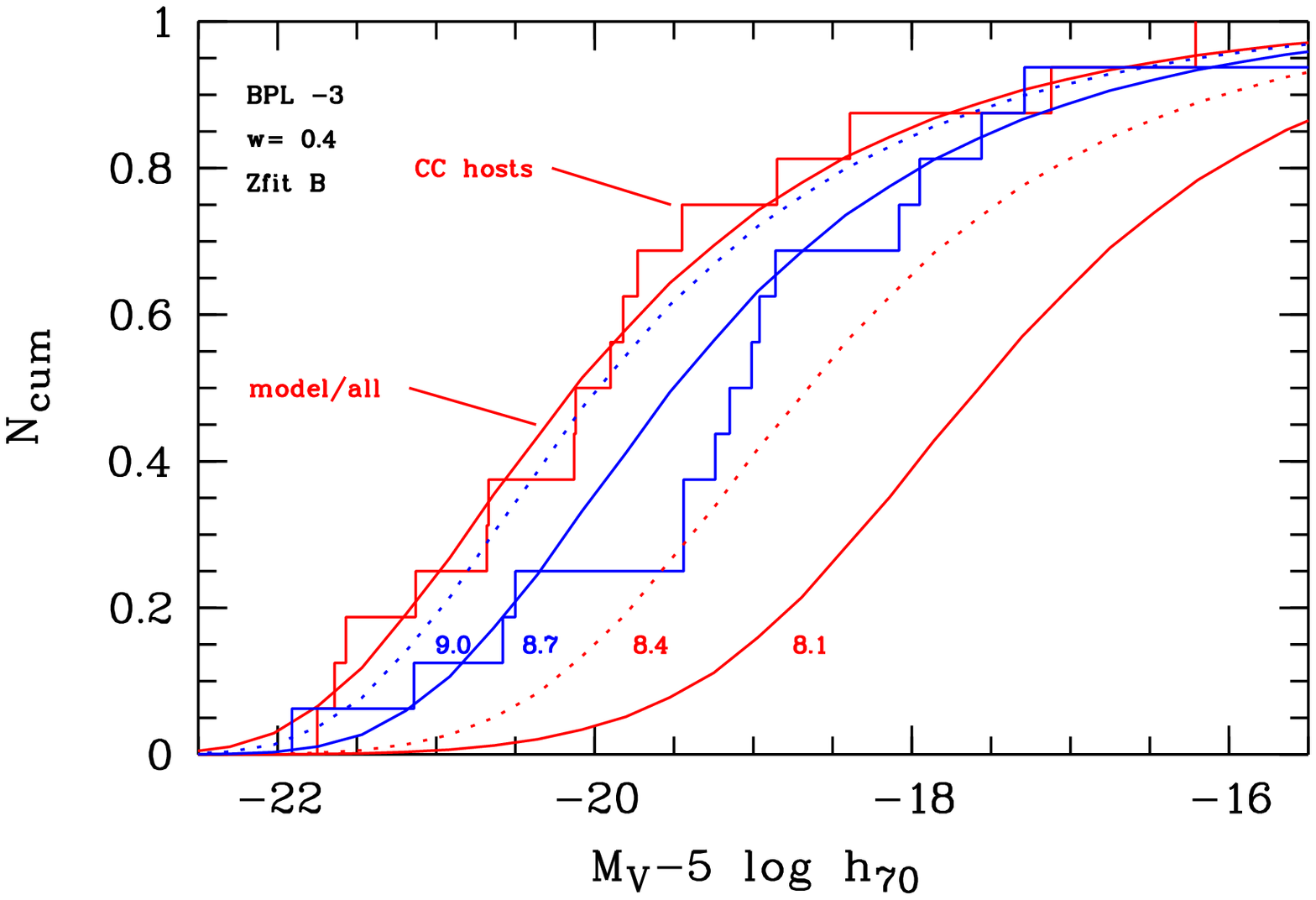}
\includegraphics[clip,angle=270,width=0.47\hsize]{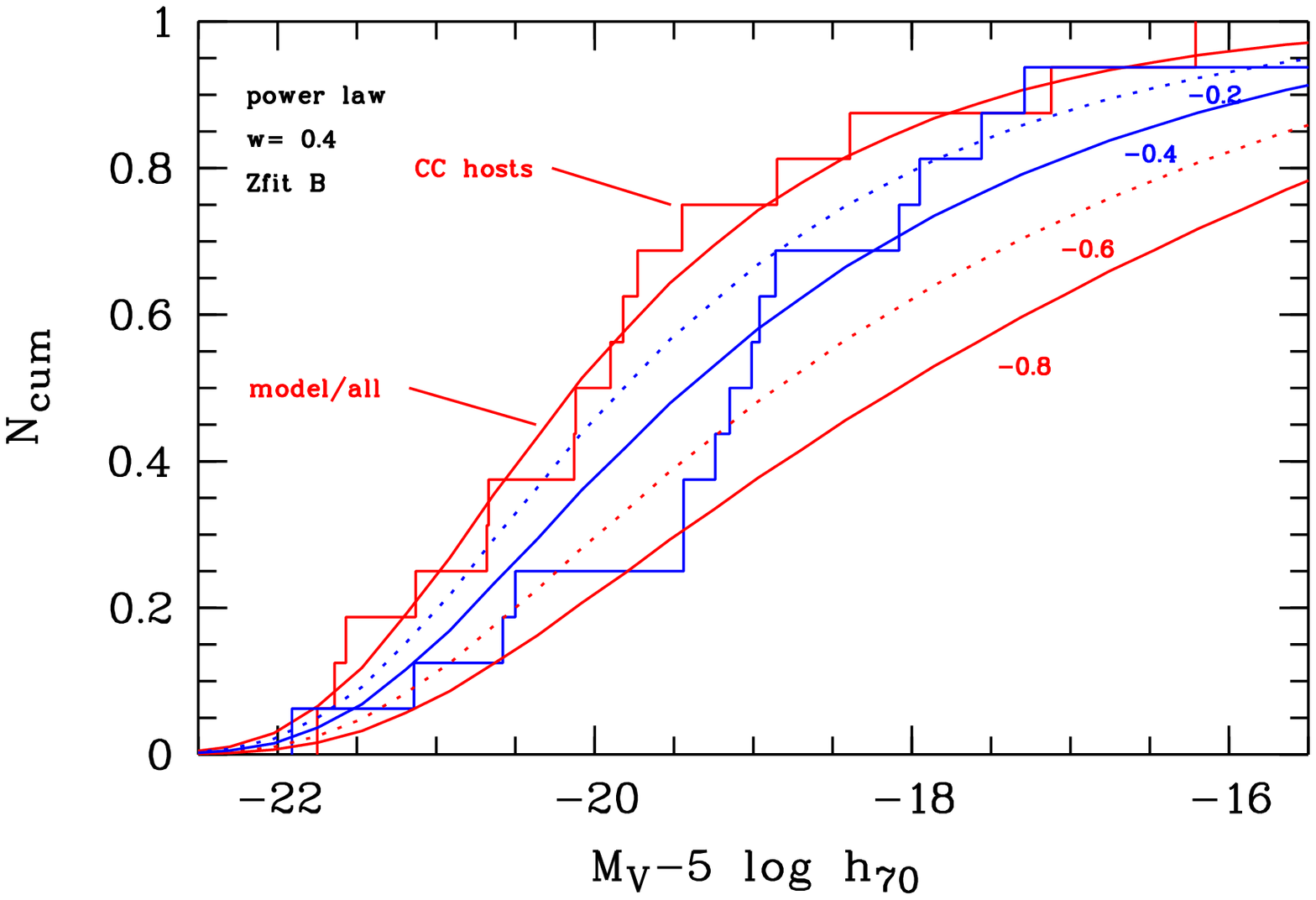}
 \hspace{5mm}
\includegraphics[clip,angle=270,width=0.47\hsize]{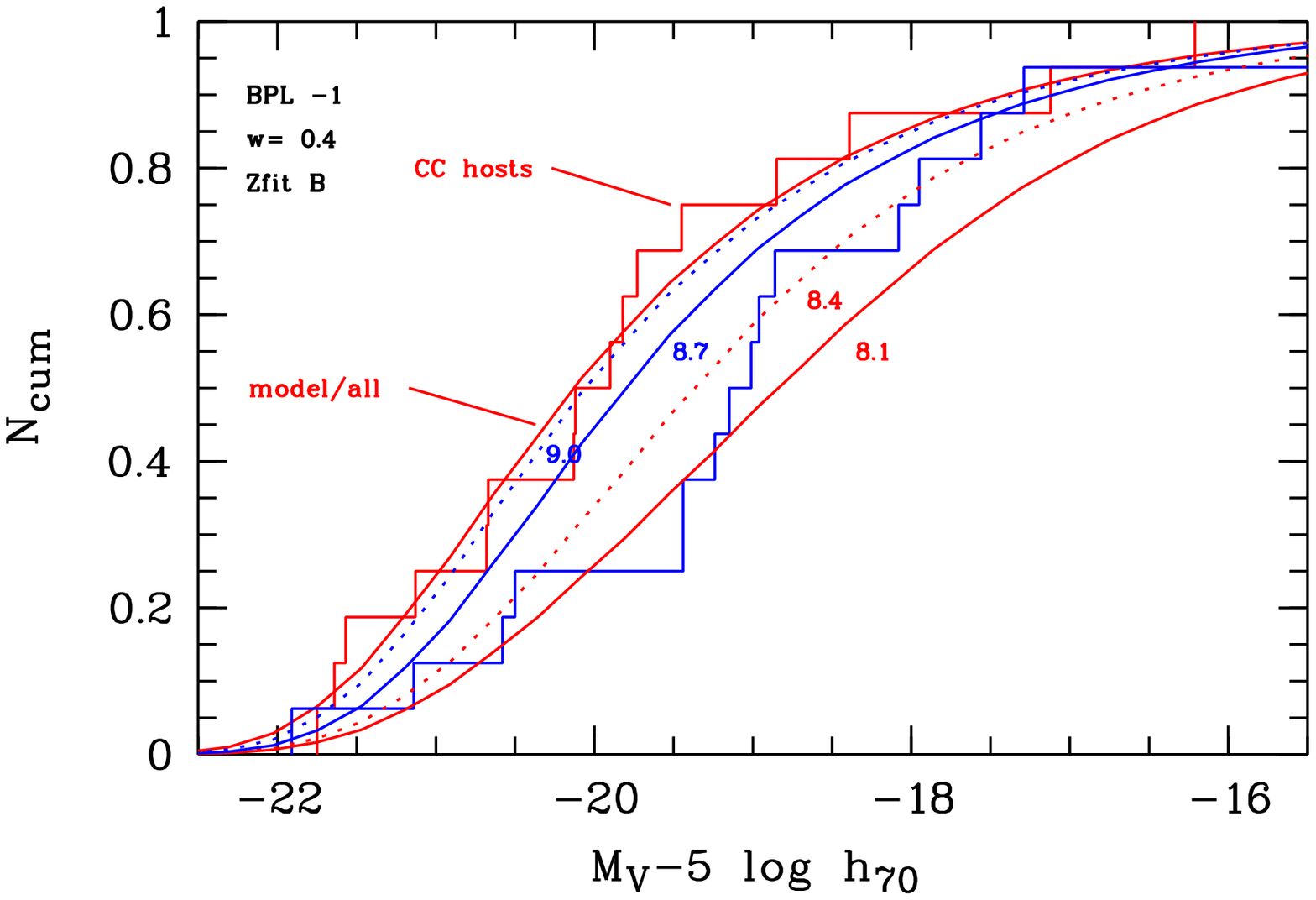}
\caption{{\it Top left:} Cutoff models for a sequence of metallicity
cutoffs at ($Z_{\rm lim} =\{9.0,8.7,8.4,8.1\}$). The reference model 
for {\it all\/} galaxies has no metallicity dependence.
{\it Bottom left:} Power-law efficiencies with $c=\{-0.2,-0.4,-0.6,
-0.8\}$. All models predict too many bursts in faint galaxies. This 
is a consequence of the divergence to infinity of the LGRB efficiency 
at low metallicity. Varying $w$ has no effect on power-law models.
{\it Right:} Broken power laws: The softer the break in the efficiency 
function, the lower the break metallicity in the best-fitting models. 
A hard cutoff ($c=-3$, {\it top}) is best fit with $Z_{\rm lim}\simeq 
8.7$ (i.e. solar), while a very soft break ($c=-1$, {\it bottom}) is 
best matched at $Z_{\rm lim} \simeq 8.2$. In the latter case the LGRB
efficiency at solar metallicity is still a third of the plateau value.
\label{LZrange}}
\end{figure*}

\subsection{Event efficiencies and their dependence on metallicity}

Here, we specify our descriptions for the dependence of the LGRB
efficiency on the metallicity of young stars. We consider two simple
one-parameter models and one two-parameter model, which is a
compromise between the first two.  Any {\em adhoc} model that is more
complicated than these makes little sense given the small number of 16
LGRBs in the sample. Our adopted prescriptions are:

\begin{itemize}
\item a sharp high-$Z_{\rm O}$ cutoff:
  \begin{equation}\label{Cut}
    \eta (Z_{\rm O}) = \left\{  \begin{array}{cl}
	1 	& \mbox{~if $Z_{\rm O}\le Z_{\rm lim}$}, \\
	0	& \mbox{~if $Z_{\rm O} > Z_{\rm lim}$}.  
	\end{array} \right.
  \end{equation}
\item a power law, which unfortunately diverges at low $Z_{\rm O}$:
  \begin{equation}\label{PL}
	\log \eta (Z_{\rm O}) = -c\, Z_{\rm O}.
  \end{equation}
\item a broken power law, which is flat as in (\ref{Cut}) below a 
break metallicity $Z_*$ and declines as in (\ref{PL}) above the break:
  \begin{equation}\label{BPL}
    \log \eta(Z_{\rm O}) = \left\{ \begin{array}{cl} 0 &
        \mbox{~if $Z_{\rm O}-Z_*\le 0$}, \\ 
     -c\, (Z_{\rm O} - Z_*) &
	\mbox{~if $Z_{\rm O}-Z_* > 0$}.
	\end{array} \right.
  \end{equation}
\end{itemize}
Note, that for the single power law no normalisation is required as we are
not dealing with absolute LGRB rates but only host distributions. Also, model
(\ref{BPL}) contains the other models as special cases, because it converges 
into (\ref{Cut}) for $c\rightarrow -\infty$, and into (\ref{PL}) for $Z_* 
\rightarrow -\infty$.

The prescriptions for event efficiencies $\eta (Z_{\rm O})$ and for the 
metallicity distributions of star-forming gas $p(Z_{\rm O}|M_V)$ in hosts
of given luminosity $M_V$ are then included into our prediction of the 
V-band host luminosity distribution, using the new integral
\begin{eqnarray*}\label{newint}
  \lefteqn{j_{280} (L_{V,\rm lim}) = dL_{280}/dL_V \times} \\
       & & \int_{L_{V,\rm lim}}^{\infty} 
	\int_{-\infty}^{\infty} p(Z_{\rm O}|L_V) \, \eta (Z_{\rm O}) \,
	 L_{280}(L_V) \, \phi(L_{V}) \, dL_V \, dZ_{\rm O} ~.
\end{eqnarray*}
Again, this figure is now proportional to the LGRB event rate expected
for the assumed efficiency function.

\section{Comparison with the data}

In this section, we compare our metallicity-dependent models with 
the host galaxy data. However, in Sect.~5.2 we will discuss empirical 
evidence that the LGRB hosts discussed here are on the L--Z relation
as far as we know, supporting the validity of our basic approach.

\subsection{Cutoff models}

Here, we use a constant efficiency up to a cutoff metallicity defined 
in eq.~(\ref{Cut}) and consider a sequence of cutoff values of $Z_{\rm
lim}=\{9.0, 8.7, 8.4, 8.1\}$, i.e. roughly twice solar, solar, half
solar and a quarter solar. The top left panel in Fig.~\ref{LZrange} 
shows the resulting models alongside the reference model with {\it 
all\/} galaxies that fits the CC SN hosts. The most likely cutoff 
metallicity can be readily obtained from the plot as $\sim 8.7$. 
Already a cutoff at 8.4 clearly underpredicts the host luminosities. 

We also investigated the dependence of this results on uncertainties 
in the $L$--$Z_{\rm O}$-relation by using the limiting fits A and C 
from Tab.~\ref{KK04fits} to define fiducial confidence intervals.
We find that changes in the slope of the $L$--$Z_{\rm O}$-relation 
translate directly into changes in the luminosity spread of the 
curves predicted for a fixed set of cutoff metallicities, which is 
expected. Fortuitously, the data lie in a range that appears to be 
not very sensitive to the choice of the $L$--$Z_{\rm O}$-fit. This
is a result of the pivotal point in the $L$--$Z_{\rm O}$-fit being
close to $M=-20$. There are not enough GRB hosts far away from the
pivotal point of the $L$--$Z_{\rm O}$-fit to let the slope make a 
large difference for the best-fitting cutoff metallicity. Instead,
the slope of the relation affects the formal errors on the cutoff, 
which one obtains from a fit to the data (see Sect.~4.4).

\subsection{Power-law efficiencies}

As an alternative to a sharp cutoff model, we now test an efficiency
function that smoothly changes with metallicity and choose a simple
power law. The reference model of a constant efficiency for all galaxies 
is always shown for comparison, and corresponds to $c=0$ in eq. (\ref{PL}). 
Fig.~\ref{LZrange} further shows the curves for power-law slopes of $-0.2$, 
$-0.4$, $-0.6$ and $-0.8$, which deviate increasingly from the reference 
model as high-$Z_{\rm O}$ GRBs are more and more suppressed relative to 
low-$Z_{\rm O}$ events.

Any variation in $w$ has no effect on the curves, because the
symmetric form of the $Z_{\rm O}$ distribution is averaged out with
a power-law efficiency.  Hence, there is no need to reproduce the curves 
for any other $w$. Furthermore, the slopes $c$ of the efficiency curves 
are degenerate with the slope of the $L$--$Z_{\rm O}$-relation. So, fits A
or C will produce the same curves, if suitable $c$-values are chosen.
All curves for power laws predict a substantial contribution from
extremely low-luminosity galaxies, which are not observed among GRB
hosts. This is due to the unrealistic divergence of the GRB efficiency
towards zero metallicity.

\subsection{Broken power laws}

The broken power law with flat low-$Z_{\rm O}$ efficiency combines the
properties of the two previous models, a gradual decline towards
higher metallicity and a constant, non-diverging efficiency at lower
metallicity. Hence, this model avoids the unrealistic and problematic
divergence of the single power law, while being less restrictive than the
cutoff model, providing the ``best of both worlds''.

The model is now described by two parameters, which provides more
freedom to fit the data, but also introduces some degeneracy. While
hard breaks are very similar to the cutoff models, the metallicity
break locations for softer breaks become increasingly less
constrained.  Especially, with only 16 GRBs the data do not provide
sufficient constraints for the model. Conversely, we conclude that
these data do not yet rule out a variety of models.

Fig.~\ref{LZrange} shows results for two broken power laws to illustrate
the trends with the slope $c$. The sequence of models shows breaks at
$Z_*=\{9.0,8.7, 8.4,8.1\}$; a constant efficiency matching the CC SN
data is also shown. We find that the data can not differentiate between 
a soft and a hard break. Instead, a hardening of the break only leads 
to higher break metallicities.  While the assumption of a hard cutoff 
suggests $Z_*\simeq 8.7$, softening the break to $c=-1$ moves $Z_*$ to 
$\sim 8.2$.

\begin{figure}
\centering
\includegraphics[clip,angle=270,width=0.8\hsize]{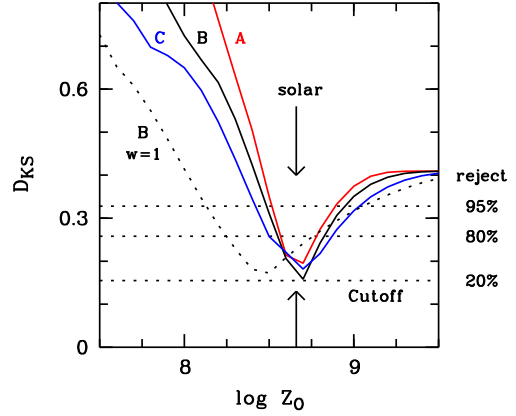}
\caption{Kolmogorov-Smirnov tests: Cutoff models with 
$w=0.4$ (solid curves) and different $L$-$Z_{\rm O}$-fits. The dashed 
curve depicts a model with extreme metallicity inhomogeneities $w=1$. 
\label{KS}}
\end{figure}

\begin{figure}
\centering
\includegraphics[clip,angle=270,width=0.8\hsize]{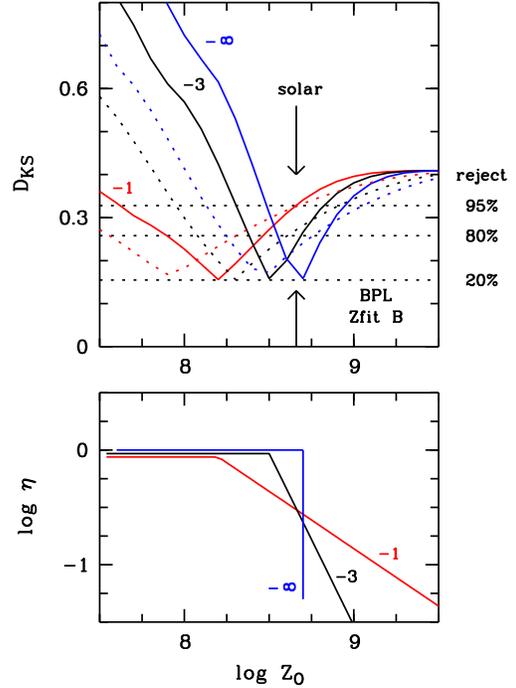}
\caption{Kolmogorov-Smirnov tests: 
{\it Top panel:} Broken power-law models for different slopes 
$c=-1$, $-3$ and $-\infty$ (the $-\infty$-case is identical to a model with
a sharp cutoff).
Dashed curves are for $w=1$ (as in Fig.~6).
{\it Bottom panel:} The efficiency function for the best-fitting 
power-law models, slightly offset vertically for clarity. \label{KS2}}
\end{figure}

\subsection{Kolmogorov-Smirnov tests and results at redshift $<1$}

In the present analysis systematic uncertainties appear to dominate,
ranging from the $L$--$Z_{\rm O}$-relation over the typical
metallicity spread in the star-forming gas of galaxies at $z\simeq
0.7$ to the form of the faint end of the ultraviolet galaxy
luminosity function. However, it is straightforward to obtain
quantitative confidence intervals on $Z_{\rm lim}$ (as shown in 
Fig.~\ref{KS}) or $Z_*$ (Fig.~\ref{KS2}) for a particular model, 
using Kolmogorov-Smirnov (KS) tests. We repeat, that the KS test 
yielded a highly acceptable value of $D_{\rm KS}=0.12$ for the 
match between the CC SNe and the reference model of `all' galaxies.

Fig.~\ref{KS} shows the results for the cutoff models. The solid curves 
show the three different $L$--$Z_{\rm O}$-relations with a realistic 
metallicity spread of $w=0.4$. All three favour a cutoff near solar 
metallicity and differ mostly in the width of the confidence interval,
which is a result of the different slopes of the relations. We obtain 
$Z_{\rm lim}=8.7\pm 0.22$ at 95\% c.l. for the most likely $L$--$Z_{
\rm O}$-relation B or $\pm 0.3$ if allowing for any relation. 

But how would these results be affected if the internal metallicity
dispersion of star-forming galaxies was truely much larger than ever
anticipated? Would this allow the low-metallicity tails in luminous
galaxies to produce GRBs and make the host data consistent with much
lower metallicity cutoffs? For illustration, we modelled an extremely
wide metallicity spread of the star-forming populations at given host
$M_V$, from 1/10th of the mean to $10\times$ the mean. While such a 
model is unrealistic, the location of the cutoff metalllicity is only
lowered by $\sim 0.25$~dex (see dashed line in Fig.~\ref{KS}).
Invoking such extreme metallicity inhomogeneities among young stellar 
populations still rules out cutoffs of $Z_{\rm lim} < 8.1 \approx
Z_\odot /4$ at 95\% confidence.

Fig.~\ref{KS2} illustrates the broken power-law models. Softening the
break, from a hard cutoff ($c=-\infty$) over $c=-3$ to a slope of $c=-1$
moves the favoured break metallicity to lower values and widens the 
confidence interval. A more gradual change in GRB efficiency leads to a 
more gradual rise of the cumulative host luminosity distribution. A soft 
break is also a less pronounced feature in the efficiency function $\eta
(Z_{\rm O})$. The best-fitting $\eta$ functions shown in the bottom 
panel illustrate that even softer breaks require the $\eta (Z_\odot) 
\approx 1/3$. 

In summary, we have explored for the $z=[0.2,1.0]$ sample the impact of 
the two most important uncertainties by comparing results for different 
sets of assumptions. Three alternative $L$--$Z_{\rm O}$-relations and 
different degrees of internal metallicity scatter in the star-forming gas 
are put in contrast in Fig.~\ref{KS}. 
Varying the $L$--$Z_{\rm O}$-relations around the pivotal points provided
by the observations of $z\simeq 0.7$ galaxies appears to have very little
effect. Changing the internal metallicity scatter of typical galaxies can
have a significant effect when assuming unconventionally extreme degrees.

We further calculated the fractions of the young stellar population that
can produce LGRB progenitors for $z\simeq 0.7$, the mean redshift of
our LGRB sample, given the various efficiency prescriptions. In
the metallicity-independent model where all galaxies contribute and
match the CC SN distribution, this fraction is defined as equal to
1. In the cutoff model the best-fitting value $Z_{\rm lim}=8.7\pm 0.3$
corresponds to fractions of $0.55\pm 0.25$. This means that because of
the metallicity constraint the LGRB production efficiency is reduced
by a factor of about one half compared to the efficiency if there were
no metallicity constraint. This number is also consistent with half of
the star formation and half of the CC SNe originating from galaxies with
lower metallicity than solar, or with lower luminosity than $M_B=-20$.
In the broken power-law models we find very similar fractions independent 
of the ambiguity between break softness and break location. In the single 
power-law model the fractions can not be normalised because the efficiency 
diverges at low $Z_{\rm O}$. However, these models did not fit the data 
for the same reason.

\subsection{High-redshift host galaxies}

The purpose of restricting the GRB sample to low redshift was to
compare it directly against the CC SN sample and to quantitatively
investigate the role of metallicity in a regime where we have ample
knowledge about the star formation and metallicity of the galaxy
population. After obtaining a possible value for a metallicity cutoff
around $Z_{\rm lim}\simeq 8.7$, we can check qualitatively whether 
this is in agreement with the observations of higher-redshift hosts.
The F06 sample contains eight host galaxies in the range $z=[1.5,3.5]$, 
half of which have luminosities brighter than $M_V=-20$. This reflects
the fact, that at higher redshift larger galaxies were more actively 
forming stars than they do at lower-redshift \citep[e.g.][]{PG05}. 

A recent study of $z\simeq 2$-galaxies \citep{Erb06} showed that 
there is a mass--metallicity relation in place but no well-defined 
$L$--$Z_{\rm O}$-relation due to strong variations in the M/L-ratios 
of galaxies. In their study, the most massive galaxies reach $Z_{\rm 
O}=8.6$, which is on a $T_e$-based calibration and probably equivalent 
to $\sim 8.9$ on the KK04 scale. A limit at $Z_{\rm O}\le 8.7$ would
correspond to a mass limit of $M_*/M_\odot \le 10^{10.5}$. In fact,
the sample contains plenty of luminous galaxies below this mass with 
$M_V$ up to $-22$. Hence, our findings from the $z<1$-sample are not
in conflict with the luminosities of higher-redshift hosts. On the 
other hand, the Erb et al. 
sample contains galaxies with metallicities above our estimated cutoff,
suggesting that at $z\simeq 2$ LGRBs are not entirely unbiased 
tracers of star formation, which they might be at yet higher redshift. 

\begin{figure}
\centering
\includegraphics[clip,angle=270,width=0.78\hsize]{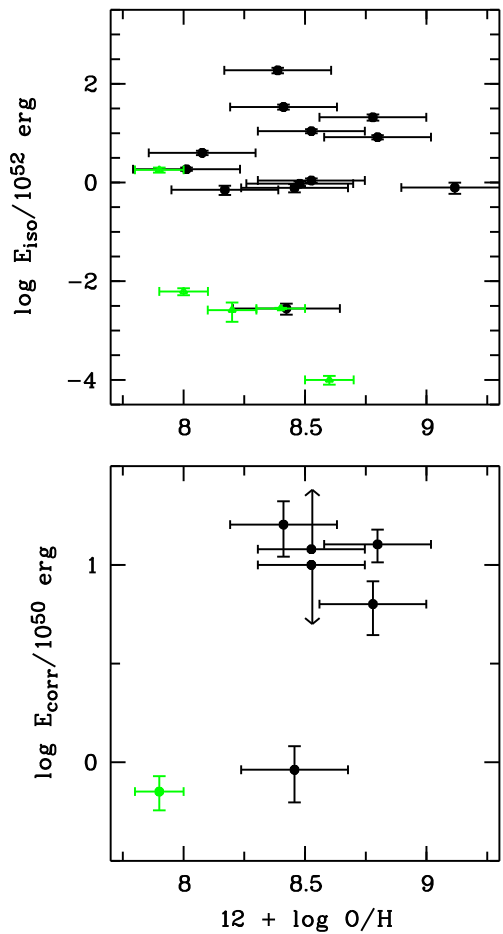}
\caption{{\it Top panel:} Comparison of the fiducial isotropic LGRB 
energy $E_{\rm iso}$ with the host metallicity, as estimated from the 
$L$--$Z_{\rm O}$-relation for $z>0.2$-bursts (black points) and as 
measured from spectroscopy for four $z<0.2$-bursts (grey points).
{\it Bottom panel:} Comparison of the LGRB energy $E_{\rm corr}$ 
corrected for beaming by \citet{GGL04} with the host metallicity.
\label{EisoZ}}
\end{figure}

\section{Discussion}

\subsection{Caveats}

One difficulty with the present analysis could be host confusion. If
the GRBs took preferentially place in low-luminosity companions to the
formally identified hosts, then the observed host luminosity
distribution would be biased to higher luminosities, suggesting higher
metallicity cutoffs. It is beyond the scope of this paper to
investigate whether HST observations could resolve such dwarfs from
their larger, mostly LMC-type companions at $z<1$.  However, if most
LGRBs occurred in such small galaxies, one would also expect a large
population of sufficiently isolated small dwarfs, where no confusion
would take place. But the small number of observed isolated hosts with 
$M_V>-17$ leaves little room for any adjustment of our results.

On the contrary, it is possible that more massive LGRB hosts are 
missing from the sample: luminous dust-obscured star-bursting galaxies 
harbour a significant fraction of the overall cosmic star formation. 
If LGRBs occur in these, their optical afterglows may be invisible,
which would depopulate the host sample specifically at the bright end.
The CC SNe host sample would be less affected by this effect, as CC
SNe have a lower average progenitor mass and hence longer average time 
delay from formation to explosion. This allows the progenitors to leave
their immediate formation environment and travel into less obscured
regions. The effect of progenitor mass on the distance to the likely 
formation site has been shown to apply in the comparison of SNe II and
SNe Ib/c \citep{JA06}. The sample analysed here lacks by definition all
so-called 'dark bursts', which may be related to massive dusty galaxies:
GRB 020127 only produced an X-ray afterglow, and the host was identified
as a $z\sim 1.9$ dusty ERO galaxy with 5~$L^*$ luminosity \citep{Be06}.
\citet{Le06} observed an extremely red ($R-K\sim 6$) afterglow to GRB
030115 in an ERO host galaxy with $R-K\sim 5$, suggesting that this
burst was highly obscured by dust. The afterglow was only identified 
with NIR observations and would have been missed by optical follow-up
alone.
Again, for the more recent GRB 050223 only an X-ray afterglow was seen
by {\it Swift}, but no optical afterglow. The host was shown to be a
dusty galaxy with $A_V>2$ \citep{Pel06}. {\it If there is a significant 
observational bias against LGRBs in massive dust-enshrouded galaxies, 
then the metallicity-dependence of LGRB rates may be rather weaker than 
our findings presently suggest.}

Uncertainties also exist with respect to the calibration of oxygen
gas-phase metallicities and on how the oxygen abundance is
representative of other elements. E.g., $\alpha$-elements are often
found to be enhanced in systems where the star-formation timescale is
much shorter than the timescale for Type Ia supernovae. Depending on
the progenitor model for LGRBs, different elements might be most
responsible for affecting opacities in stellar atmospheres/winds and
hence stellar evolution.  However, $\alpha$-enhancements are typically
observed in old ellipticals which are not at all common LGRB hosts or
typical star-forming galaxies, and thus should not affect our analysis
appreciably.

In summary, solar metallicity appears to mark a pivotal point of
roughly constant efficiency apparently independent of the softness of
the break.  At $z\sim 0.7$, this metallicity seems typical for
galaxies of $M_B-5 \log h_{70}=-20$ in either $L$--$Z_{\rm
O}$-relation. The number ratio of CC SNe to either side of this host
luminosity cut is 1:1, whereas it is 1:3 for LGRBs. If there was any
doubt about the calibration of the oxygen abundance at $z\sim 0.7$, or
the relevance of oxygen in comparison to other elements, then we could
rephrase our main conclusion such that a most likely cutoff for LGRBs
is around the mean metallicity (by any measurement) of $M_B=-20$
galaxies at $z\sim 0.7$.

\subsection{Are GRB hosts on the luminosity-metallicity relation?}

Several recent studies have aimed at constraining LGRB progenitors
from direct measurements of host metallicities \citep{So05,Sav06,Sta06} 
using very small samples:  \citet{So05} present measured metallicities 
for three LGRB hosts with values of $\sim 8.2\ldots 8.7$. 
\citet{Sav06} consider seven hosts at $z=[0.4,1.0]$ and find mean galaxy
metallicities of 8.3 to 8.55 for five of them, while two hosts are
clearly super-solar. They furthermore claim, that the hosts fall right
onto the regular mass--metallicity relation of normal star-forming
galaxies.  This host sample with measured metallicities lacks the
lowest-luminosity hosts and so contains preferentially higher
metallicity galaxies. Hence, their median metallicity is not
representative, but the statement on consistency with the
mass--metallicity relation is highly relevant.

\citet{Sav06b} give a number of host metallicity determinations, which
are collected from the literature and converted onto the same (KK04) 
calibrator, which is also the calibrator used in this paper for the 
$L$--$Z_{\rm O}$-relation. Five of the above hosts are actually part
of the LGRB host sample used here. Their spectroscopic metallicities
are compared to the estimated values we have assigned to them using
$L$--$Z_{\rm O}$-fit B. We find them all to be within $\pm 1/3$~dex
of the relation with a mean of $-0.034$~dex, which is consistent with 
no bias within the statistical power of five objects.

This results is not surprising. A physical model for GRB progenitors
may incorporate an explicit metallicity dependence, but we can not
expect the progenitor to know explicitely about galaxy parameters
such as luminosity or morphology. At a fixed metallicity level, we 
would then expect bursts predominantly from 
those galaxy mass bins containing the highest star-formation rate. 
Here, both the shape of the mass function and the SFR trends with 
mass play a role. Presumably, the declining mass function and the 
increasing SFR with mass cancel to some extent. 
KK04 searched for trends of galaxy 
properties within the scatter of their $L$--$Z_{\rm O}$-relation.
While they found no such trends, they particularly did NOT find
strongly star-forming galaxies on the bright or metal-poor side
of the relation. However, among very low-mass dwarf galaxies it is
conceivable, that strong variations in SFR with time lead to such
strong variations in mass-to-light ratio, such that CC SNe and 
LGRBs are found predominantly on the bright side of the relation.
Presumably, this would have no effect for our analysis, as the mean
metallicity bias would be small and remain within the flat portion
of the parametric efficiency functions considered in this paper.

\subsection{Global vs. local metallicity measurements}

We would like to point out, that even spectroscopic, but spatially 
unresolved, measurements of a mean host metallicity do not directly 
reflect the metallicity of the young stellar population near the LGRB 
progenitor. In fact, the error in the progenitor metallicity is
dominated by the internal $Z_{\rm O}$ dispersion of the galaxy, which 
is estimated to be as large as the dispersion of galaxy-averaged 
$Z_{\rm O}$ values in the $L$--$Z_{\rm O}$-relation ($\la 0.2$~dex). 

If we make no prior assumptions on LGRB hosts and accept the result 
that LGRB hosts follow the usual $L$--$Z_{\rm O}$-relation, then we 
can in turn estimate even inidividual host metallicities from this 
relation. These individual estimates then have errors on the order of 
the dispersion in the $L$--$Z_{\rm O}$-relation, i.e. $\la 0.2$~dex.

In other words, owing to the internal metallicity dispersion, our
indirect estimate of the progenitor metallicity via the $L$--$Z_{\rm
O}$-relation of the host galaxy should be almost as useful as an
unresolved spectroscopic observation, with an error that is roughly
larger by a factor of $\sqrt{2}$. If the internal metallicity
dispersions exceeded 0.2~dex, then the determination of integrated
metallicities of individual LGRB hosts would have very little value,
because it could as well be estimated from an $L$--$Z_{\rm
O}$-relation, while the error was mostly internal in origin.

Therefore, major progress can mostly be expected from spatially 
resolved metallicity measurements that focus on the immediate burst
environment. However, such observations are currently not possible
for GRBs at cosmological distances.

\subsection{A metallicity--LGRB energy relation?}

\citet{Sta06} suggest a tentative relation between host metallicity 
$Z_{\rm O}$ and the isotropic energy associated with the LGRB $E_{\rm 
iso}$, which they interpret as a metallicity cutoff around 0.15 solar 
for regular cosmological bursts. This conjecture is based on one 
object out of five, where a high LGRB energy coincides with low host 
metallicity.While this object was assigned a much higher host 
metallicity in a previous analysis \citep{So05}, the mass of the 
galaxy makes that interpretation very unlikely. 
While the claimed relation relies on the significance of this single 
object, it is unclear whether the other (underluminous) LGRBs are 
representative of the more energetic and distant cosmological bursts.
GRB 060218, e.g., could not even have been detected by {\it Swift} at 
redshifts $z>0.05$.

Even in the absence of spectroscopic metallicity measurements for the 
individual hosts, we can tentatively investigate such a relation for 
the full sample with known $E_{\rm iso}$ and host luminosities using
the $L$--$Z_{\rm O}$-relation.
Fig.~\ref{EisoZ} shows the result for 13 GRBs from our sample (black
data points) which have $E_{\rm iso}$ values in the literature \citep{
Am06}. We also add the five local objects discussed by \citet{Sta06}
based on their actual host $Z_{\rm O}$-data as grey data points.  At
least for the cosmological bursts (black) no relation is apparent.
However, due to the expected significant variations in jet geometry
and viewing angle among the bursts, we would not expect the fiducial
isotropic energy estimate to show very clear trends with other
parameters. \citet{GGL04} present LGRB energies $E_{\rm corr}$ that
are corrected for jet beaming effects. Correspondingly, their
Ghirlanda relation $E_{\rm corr}$ vs.  $E_{\rm peak}$ is much more
clearly defined than the original Amati relation $E_{\rm iso}$ vs.\ 
$E_{\rm peak}$ \citep{Am02}, where $E_{\rm peak}$ is the restframe photon
energy at the peak in the $\nu F_\nu$ spectrum.  In the bottom panel
of Fig.~\ref{EisoZ}, we show $E_{\rm corr}$ vs.\ the host metallicity
for the bursts with available data. We find no trend in these data.

\subsection{Theoretical implications}

From a theoretical point of view, it is not surprising that
metallicity plays an important role for LGRB progenitors. In one of
the most promising progenitor scenarios, the collapsar model (Woosley
1993; MacFadyen \& Woosley 1999), the progenitor is the rapidly
rotating core of a massive star. On the other hand, the core of a
massive star is believed to efficiently lose angular momentum during
its evolution: both within the star by hydrodynamical (e.g. Heger,
Langer \& Woosley 2000) and magnetohydrodynamical processes
(e.g. Spruit 2002) and subsequently from the surface of the star in
the form of a stellar wind. As a consequence, it seems unlikely that
most massive stars will still have cores at the time of core collapse
that rotate as rapidly as is required in the collapsar model (Heger,
Woosley \& Spruit 2005).

Indeed, this is also not necessary, since it is clear that LGRBs are
rare events, associated with less than 1 in 1000 core-collapse
supernovae (Podsiadlowski et al.\ 2004b).  One of the key unresolved
questions is what are the special circumstances that produce LGRB
progenitors: are these special conditions in single stars (e.g. Yoon
\& Langer 2005; Woosley \& Heger 2006) or do they require particular
binary channels (e.g. Fryer \& Woosley 1998; Izzard, Ramirez-Ruiz \&
Tout 2003; Fryer \& Heger 2005; Petrovic et al.\ 2005; Detmers et al.\
2006; Fryer et al.\ 2006; Podsiadlowski et al.\ 2006)? One of the
major problems in many of these models is that mass loss, both in the
red-supergiant phase and, in particular, in the Wolf-Rayet phase,
leads to very efficient angular-momentum loss. Since these wind
mass-loss rates are dependent on the metallicity (typically $\dot{M}
\propto Z^{0.5-0.7}$; e.g. Vink \& de Koter 2005), massive stars with
lower metallicity are more likely to have rapidly rotating cores at
the time of core collapse. This is a generic advantage for many of the
proposed models, both single and binary.

Specifically, Yoon \& Langer (2005) and Woosley \& Heger (2006)
recently proposed that a low-metallicity, very rapidly rotating star
may evolve essentially homogeneously during its early evolutionary
phases and may avoid a red-supergiant phase altogether, in which most
of the angular-momentum loss from the core tends to take place. As a
consequence, these stars would still preserve rapidly rotating cores
at the time of collapse, fulfilling one of the key conditions in the
collapsar model. Similarly, in some of the most promising binary
scenarios, in which two massive stars (almost) merge or interact
tidally, it is advantageous or even necessary that this occurs late in
the evolution of one of the stars, i.e. after helium core burning
(so-called case C mass transfer). The range of masses for which case C
mass transfer occurs and allows the formation of a black hole is a
strong, non-linear function of metallicity (Justham et al.\ 2006),
again favouring low-metallicity progenitors.

As this discussion illustrates, the dependence of the LGRB rate on
host metallicity can provide an important constraint on the various
proposed progenitor models. Indeed, \citet{HMM05} and \citet{YL06} have
estimated that, when the effects of magnetic fields are included,
the single, low-metallicity, high-rotation model \citep{YL05,WH06} 
requires a metallicity less than 1/5th solar. This already appears 
inconsistent with the constraints derived in this paper, which seem 
to rule out any models that {\em require} a metallicity significantly 
less than 1/2 solar.

The ongoing metallicity {\it calibration debate} might render our limit 
to lie $\sim 0.3$~dex lower, if $T_e$-based methods were correct. There 
are now explorations of a third method for measuring metallicities, the
O~{\sc II}$_{RL}$ method \citep{Pe06}, which is aimed at overcoming the 
shortcomings of both previous methods. Early results indicate that this 
method gives values in between the two debated versions, but closer to
the $R_{23}$ method. This would leave our results largely unchanged.

\section*{acknowledgements}
CW was supported by a PPARC Advanced Fellowship and appreciates the
hospitality of the IAA in Granada, where this paper was written up.
We acknowledge discussions with Andy Fruchter, Javier Gorosabel, 
Stephen Justham, Norbert Langer and Sung-Chul Yoon, as well as regular 
inspiration from the Stellar Coffee Group at Oxford. We thank the
anonymous referee for his support in improving the clarity of the 
manuscript.


\begin{thebibliography}{}

\bibitem[\protect\citeauthoryear{Allende Prieto, Lambert \& Asplund}{2001}]{ALA01}
 Allende Prieto, C., Lambert, D. L. \& Asplund, M., 2001, ApJ, 556, L63
\bibitem[\protect\citeauthoryear{Amati}{2006}]{Am06}
 Amati, L., 2006, MNRAS, 372, 233 
\bibitem[\protect\citeauthoryear{Amati et al.}{2002}]{Am02}
 Amati, L., et al., 2002, A\&A, 390, 81
\bibitem[\protect\citeauthoryear{Asplund, Grevesse \& Sauval}{2005}]{A05}
 Asplund, M., Grevesse, N. \& Sauval, A. J. 2005, in {Cosmic Abundances as 
 Records of Stellar Evolution and Nucleosynthesis}. ASP Conf. Ser. 336, p. 25. 
\bibitem[\protect\citeauthoryear{Berger et al.}{2006}]{Be06}
 Berger, E., Fox, D. B., Kulkarni, S. R., Frail, D. A. \& Djorgovski, S. G.,
 2006, ApJ, submitted, astro-ph/0609170
\bibitem[\protect\citeauthoryear{Brown, Lee \& Bethe}{1999}]{B99}
 Brown, G. E., Lee, C.-H. \& Bethe, H. A. 1999, New Astronomy, 4, 313
\bibitem[\protect\citeauthoryear{Brown et al.}{2001}]{B01}
 Brown, G. E., Heger, A., Langer, N., Lee, C.-H., Wellstein, S., \&
 Bethe, H. 2001, New Astronomy, 6, 457
\bibitem[\protect\citeauthoryear{Caldwell et al.}{2006}]{C06}
 Caldwell, J. A. R., McIntosh, D. H., Rix, H.-W. et al., 2006, ApJ, in press,
 astro-ph/0510782
\bibitem[\protect\citeauthoryear{Christensen, Hjorth \& Gorosabel}{2004}]{CHG04}
 Christensen, L., Hjorth, J. \& Gorosabel, J., 2004, A\&A, 425, 913
\bibitem[\protect\citeauthoryear{Dahlen et al.}{2004}]{Da04}
 Dahlen, T., Strolger, L.-G., Riess, A. G. et al., 2004, ApJ, 613, 189
\bibitem[\protect\citeauthoryear{Detmers, Langer \& Podsiadlowski}{2006}]{DLP06}
 Detmers, R., Langer, N., \& Podsiadlowski, Ph. 2006, in preparation
\bibitem[\protect\citeauthoryear{Eldridge \& Tout}{2004}]{ET04}
 Eldridge, J. J. \& Tout, C. A. 2004, MNRAS, 253, 86
\bibitem[\protect\citeauthoryear{Erb et al.}{2006}]{Erb06}
 Erb, D. K., Shapley, A. E., Pettini, M., Steidel, C. C., Reddy, N. A.,
 Adelberger, K. L., 2006, ApJ, 644, 813
\bibitem[\protect\citeauthoryear{Fruchter et al.}{1999}]{F99}
 Fruchter, A. S., et al., 1999, ApJ, 519, 13
\bibitem[\protect\citeauthoryear{Fruchter et al.}{2006}]{F06}
 Fruchter, A. S., et al., 2006, Nature, 441, 463
\bibitem[\protect\citeauthoryear{Fryer \& Heger}{2005}]{FH05}
 Fryer, C. L, \& Heger, A. 2005, ApJ, 623, 302
\bibitem[\protect\citeauthoryear{Fryer, Rockefeller \& Young}{2006}]{FRY06}
 Fryer, C. L., Rockefeller, G., \& Young, P. A. 2006, ApJ, 647, 1269
\bibitem[\protect\citeauthoryear{Fryer \& Woosley}{1998}]{FW98}
 Fryer, C. L., \& Woosley, S. E. 1998, ApJ, 502, L9 
\bibitem[\protect\citeauthoryear{Ghirlanda, Ghisellini \& Lazzati}{2004}]{GGL04}
 Ghirlanda, G., Ghisellini, G. \& Lazzati, D., 2004, ApJ, 616, 331
\bibitem[\protect\citeauthoryear{Giavalisco et al.}{2004}]{G04}
 Giavalisco, M. et al., 2004, ApJ, 600, L93
\bibitem[\protect\citeauthoryear{Han, Podsiadlowski \& Eggleton}{1995}]{H95}
 Han, Z., Podsiadlowski, Ph., \& Eggleton, P. P., 1995, MNRAS, 272, 800
\bibitem[\protect\citeauthoryear{Heger et al.}{2003}]{H03}
 Heger, A., Fryer, C. L., Woosley, S. E., Langer, N., \& Hartmann, D. H.
 2003, ApJ, 591, 288
\bibitem[\protect\citeauthoryear{Heger, Langer \& Woosley}{2000}]{HLW00}
 Heger, A., Langer, N., \& Woosley, S. E. 2000, ApJ, 528, 368
\bibitem[\protect\citeauthoryear{Heger, Woosley \& Spruit}{2005}]{HWS05}
 Heger, A., Woosley, S. E., \& Spruit, H. C. 2005, ApJ, 626, 350
 \bibitem[\protect\citeauthoryear{Hirschi, Meynet \& Maeder}{2005}]{HMM05}
 Hirschi, R., Meynet, G. \& Maeder, A., 2005, A\&A, 443, 581
\bibitem[\protect\citeauthoryear{Izzard, Ramirez-Ruiz \& Tout}{2004}]{IRT04}
 Izzard, R. G., Ramirez-Ruiz, E., \& Tout, C. A. 2004, MNRAS, 348, 1215
\bibitem[\protect\citeauthoryear{Jakobsson et al.}{2006}]{J06}
 Jakobsson, P., Levan, A., et al., 2006, A\&A, 447, 897
\bibitem[\protect\citeauthoryear{James \& Anderson}{2006}]{JA06}
 James, P. A. \& Anderson, J. P., 2006, A\&A, 453, 57
\bibitem[\protect\citeauthoryear{Justham, Podsiadlowski \& Rappaport}{2006}]{JPR06}
 Justham, S., Podsiadlowski, Ph., \& Rappaport, S. 2006, in preparation
\bibitem[\protect\citeauthoryear{Kennicutt}{1998}]{K98}
 Kennicutt, R. C., 1998, ARA\&A, 36, 189
\bibitem[\protect\citeauthoryear{Kennicutt et al.}{2003}]{K03}
 Kennicutt, R. C., Bresolin, F. \& Garnett, D. R., 2003, ApJ, 591, 801
\bibitem[\protect\citeauthoryear{Kobulnicky \& Kewley}{2004}]{KK04}
 Kobulnicky, H. A. \& Kewley, L. J., 2004, ApJ, 617, 240
\bibitem[\protect\citeauthoryear{Klose et al.}{2003}]{Kl03}
 Klose, S., Henden, A. A., Greiner, J., Hartmann, D. H., et al., 2003,
 ApJ, 592, 1025
\bibitem[\protect\citeauthoryear{Lee et al.}{2006}]{Lee06}
 Lee, H., Skillman, E. D. et al., 2006, ApJ, 647, 970
\bibitem[\protect\citeauthoryear{Le Floc'h et al.}{2003}]{LeF03}
 Le Floc'h, E., Duc, P.-A., Mirabel, I. F., Sanders, D. B. et al.,
 2003, A\&A, 400, 499
\bibitem[\protect\citeauthoryear{Le Floc'h et al.}{2006}]{L06}
 Le Floc'h, E., Charmandaris, V., Forrest, W. J., Mirabel, I. F., Armus, L.
 Devost, D., 2006, ApJ, 642, 636
\bibitem[\protect\citeauthoryear{Levan et al.}{2006}]{Le06}
 Levan, A., Fruchter, A., Rhoads, J., Mobasher, B., Tanvir, N. et al.,
 2006, ApJ, 647, 471 
\bibitem[\protect\citeauthoryear{Liang et al.}{2006}]{Li06}
 Liang, E., Zhang, B., Virgili, F. \& Dai, Z. G., 2006, ApJ, submitted,
 (astro-ph/0605200)
\bibitem[\protect\citeauthoryear{Mannucci et al.}{2003}]{Man03}
 Mannucci, F., Maiolino. R., Cresci, G., Della Valle, M., Vanzi, L. et al.,
 2003, A\&A, 401, 519
\bibitem[\protect\citeauthoryear{MacFayden \& Woosley}{1999}]{MW99}
 MacFayden, A. I. \& Woosley, S. E., 1999, ApJ, 524, 262
\bibitem[\protect\citeauthoryear{Peimbert et al.}{2006}]{Pe06}
 Peimbert, M., Peimbert, A., Esteban, C., Garcia-Rojas, J., et al., 
 2006, in {\it First light science with the Gran Telescopio Canarias},
 RevMexAA(SC), astro-ph/0608440
\bibitem[\protect\citeauthoryear{Pellizza et al.}{2006}]{Pel06}
 Pellizza, L. J., Duc, P.-A., Le Floc'h, E., Mirabel, I. F., et al., 
 2006, A\&A, 459, L5
\bibitem[\protect\citeauthoryear{Perez-Gonzalez et al.}{2005}]{PG05}
 Perez-Gonzalez, P. G., Rieke, G. H., et al. 2005, ApJ, 630, 82
\bibitem[\protect\citeauthoryear{Petrovic et al.}{2004}]{PLYH04}
 Petrovic, J., Langer, N., Yoon, S.-C., \& Heger, A. 2004, A\&A, 435, 247
\bibitem[\protect\citeauthoryear{Pfahl et al.}{2002}]{Pf02}
 Pfahl, E., Rappaport, S., Podsiadlowski, Ph., \& Spruit, H. 2002, ApJ, 574, 364
\bibitem[\protect\citeauthoryear{Podsiadlowski et al.}{2004a}]{P04a}
 Podsiadlowski, Ph., Langer, N., Poelarends, A. J. T., Rappaport, S., 
 Heger, A., \& Pfahl, E. 2004b, ApJ, 612, 1044
\bibitem[\protect\citeauthoryear{Podsiadlowski et al.}{2004b}]{P04b}
 Podsiadlowski, Ph., Mazzali, P. A., Nomoto, K., Lazzati, D., \& Cappellaro, E.,
 2004a, ApJ, 607, L17
\bibitem[\protect\citeauthoryear{Podsiadlowski et al.}{2006}]{PIJR06}
 Podsiadlowski, Ph., Ivanova, N., Justham, S., \& Rappaport, S. 2006,
 in preparation
\bibitem[\protect\citeauthoryear{Rix et al.}{2004}]{R04}
 Rix, H.-W., et al., 2004, ApJS, 152, 163
\bibitem[\protect\citeauthoryear{Russell \& Bessell}{1989}]{RB89}
 Russell, S. C. \& Bessell, M. S. 1989, ApJS, 70, 865
\bibitem[\protect\citeauthoryear{Russell \& Dopita}{1990}]{RD90}
 Russell, S. C. \& Dopita, M. A. 1990, ApJS, 74, 93 
\bibitem[\protect\citeauthoryear{Savaglio et al.}{2005}]{Sav05}
 Savaglio, S., Glazebrook, K., Le Borgne, D., et al. 2005, ApJ, 635, 260
\bibitem[\protect\citeauthoryear{Savaglio, Glazebrook \& Le Borgne}{2006}]{Sav06}
 Savaglio, S., Glazebrook, K. \& Le Borgne, D., 2006, in {\it Gamma-ray bursts
 in the Swift era}, eds. S. Holt, N. Gehrels \& J. Nousek. AIP Conf. Proc.,
 Vol. 836, p. 540-545
\bibitem[\protect\citeauthoryear{Savaglio}{2006}]{Sav06b}
 Savaglio, S., 2006, New Journal of Physics, 8, 195
\bibitem[\protect\citeauthoryear{Soderberg et al.}{2006}]{Sod06}
 Soderberg, A. M., Kulkarni, S. R., Nakar, E. et al., 2006, Nature, 442, 1014
\bibitem[\protect\citeauthoryear{Sollerman et al.}{2005}]{So05}
 Sollerman, J., \"Ostlin, G., Fynbo, J. P. U., Hjorth, J., Fruchter, A.,
 Pedersen, K., 2005, New Astronomy, 11, 103
\bibitem[\protect\citeauthoryear{Spruit}{2002}]{S02}
 Spruit, H. C. 2002, A\&A, 381, 923 
\bibitem[\protect\citeauthoryear{Stanek et al.}{2006}]{Sta06}
 Stanek, K. Z., Gnedin, O. Y., et al. 2006, ApJ, submitted, astro-ph/0604113
\bibitem[\protect\citeauthoryear{Strolger et al.}{2004}]{S04}
 Strolger, L.-G., Riess, A. G. et al., 2004, ApJ, 613, 200
\bibitem[\protect\citeauthoryear{Tremonti et al.}{2004}]{Tre04}
 Tremonti, C. A., Heckman, T. M., Kauffmann, G. et al. 2004, MNRAS, 353, 713 
\bibitem[\protect\citeauthoryear{van Paradijs et al.}{1997}]{vP97}
 van Paradijs, J., Groot, P. J., Galama, T., Kouveliotou, C., Strom, R. G.,
 et al., 1997, Nature, 386, 686
\bibitem[\protect\citeauthoryear{Vink \& de Koter}{2005}]{VdK05}
 Vink, J. S., \& de Koter, A., 2005, A\&A, 442, 587
\bibitem[\protect\citeauthoryear{Wolf et al.}{2004}]{W04}
 Wolf, C., Meisenheimer, K., Kleinheinrich, M. et al., 2004, A\&A, 421, 913
\bibitem[\protect\citeauthoryear{Wolf et al.}{2005}]{W05}
 Wolf, C., Bell, E. F., McIntosh, D. H., Rix, H.-W. et al., 2005, ApJ, 630, 771
\bibitem[\protect\citeauthoryear{Woosley}{1993}]{W93}
 Woosley, S. E., 1993, ApJ, 405, 273 
\bibitem[\protect\citeauthoryear{Woosley \& Heger}{2006}]{WH06}
 Woosley, S. E. \& Heger, A. 2006, ApJ, 637, 914
\bibitem[\protect\citeauthoryear{Yoon \& Langer}{2005}]{YL05}
 Yoon, S.-C., \& Langer, N., 2005, A\&A, 443, 643
\bibitem[\protect\citeauthoryear{Yoon, Langer \& Norman}{2006}]{YL06}
 Yoon, S.-C., Langer, N., \& Norman, C. 2006, A\&A, in press
(astro-ph/0606637)
\end{thebibliography}
\end{document}